\documentclass{PoS}

\def\asec{\ifmmode ^{\prime\prime}\else$^{\prime\prime}$\fi}

\def\degs{\ifmmode ^{\circ}\else$^{\circ}$\fi}
\def\amin{\ifmmode ^{\prime}\else$^{\prime}$\fi}
\def\asec{\ifmmode ^{\prime\prime}\else$^{\prime\prime}$\fi}
\def\degs{\ifmmode ^{\circ}\else$^{\circ}$\fi}
\def\amin{\ifmmode ^{\prime}\else$^{\prime}$\fi}

\def\cm{\mbox{\,cm}}

\def\cm3{\mbox{\,cm$^{-3}$}}

\def\lsim{\!\!\!\phantom{\le}\smash{\buildrel{}\over
 {\lower2.5dd\hbox{$\buildrel{\lower2dd\hbox{$\displaystyle<$}}\over
                                 \sim$}}}\,\,}
\def\gsim{\!\!\!\phantom{\ge}\smash{\buildrel{}\over
{\lower2.5dd\hbox{$\buildrel{\lower2dd\hbox{$\displaystyle>$}}\over
                               \sim$}}}\,\,}
\title{Unveiling the dominant gas heating mechanism in local LIRGs and ULIRGs}

\ShortTitle{Unveiling the dominant gas heating mechanism in local LIRGs and ULIRGs}

\author{\speaker{Miguel A. P\'erez-Torres}, Antonio Alberdi, Cristina Romero-Ca\~nizales \\
        Instituto de Astrofísica de Andalucía - CSIC, 18080 Granada, Spain\\
        E-mail: \email{torres@iaa.es, antxon@iaa.es, cromero@iaa.es} 
               }

\author{Luis Colina \\
Instituto de Estructura de la Materia - CSIC, 28006 Madrid, Spain\\
E-mail: \email{coina@daemir.iem.csic.es} 
}

\author{Marco Bondi, Marcello Giroletti
\\
Istituto di Radioastronomía - INAF, 40129 Bologna, Italy\\
E-mail: \email{bondi@ira.inaf.it, giroletti@ira.inaf.it} 
}

\author{Jos\'e Mar\'ia  Torrelles \\
Institut de Ci\`encies de l'Espai - CSIC, 08028 Barcelona, Spain\\
E-mail: \email{torrelles@ieec.fcr.es}
}

\author{Antonis Polatidis
\\
Joint Institute for VLBI in Europe (JIVE), 7990 AA Dwingeloo, The Netherlands\\    
E-mail: \email{polatidis@jive.nl}

}

\abstract{ We show preliminary results from a sample of Luminous and
  Ultra-Luminous Infrared Galaxies (LIRGs and ULIRGs, respectively) in
  the local universe, obtained from observations using the Very Large
  Array (VLA), the Multi-Element Radio Link Interferometer Network
  (MERLIN), and the European VLBI Network (EVN).  The main goal of our
  high-resolution, high-sensitivity radio observations is \textsl{
    to unveil the dominant gas heating mechanism in the central
    regions of local (U)LIRGs.} The main tracer of recent
  star-formation in (U)LIRGs is the explosion of core-collapse
  supernovae (CCSNe), which are the endproducts of the explosion of
  massive stars and yield bright radio events. Therefore, our
  observations will not only allow us to answer the question of the
  dominant heating mechanism in (U)LIRGs, but will yield also the CCSN
  rate and the star-formation rate (SFR) for the galaxies of the sample.}

\FullConference{The 9th European VLBI Network Symposium on The role of
  VLBI in the Golden Age for Radio Astronomy and EVN Users
  Meeting\\ September 23-26, 2008\\ Bologna, Italy}

\begin{document}

\section{Introduction}

Galaxies with infrared luminosities ($L{_{\rm IR}}[8-1000~{\mu}m] \geq
10^{11} L{_\odot}$; LIRGs), become the dominant population of
extragalactic objects in the local Universe (z $\lsim$ 0.3).  The
trigger for the intense infrared emission appears to be the strong
interaction/merger of molecular gas-rich spirals.  Galaxies at the
highest infrared luminosities ($L{_{\rm IR}}[8-1000~{\mu}m] \geq
10^{12} L{_\odot}$), known as Ultra-Luminous Infrared Galaxies
(ULIRGs), appear to be advanced mergers, and may represent an
important stage in th formation of quasi-stellar objects and powerful
radio galaxies \cite{sanders96}.  The critical question concerning
these galaxies is whether the dust in the central regions ($r
\lsim$1~kpc) is heated by a starburst or an active galactic nucleus
(AGN), or a combination of both. Mid-infrared spectroscopic studies of
ULIRGs by \cite{genzel98} suggest that the vast majority of these
galaxies are powered predominantly by recently formed massive stars,
with a significant heating from the AGN only in the most luminous
objects \cite{vei99}. These authors also found that at least half of
ULIRGs are probably powered by both an AGN and a starburst in a
circumnuclear disk or ring, which are located at typical radii
$r\simeq$700~pc from the nucleus of the galaxy, and also contain large
quantities of dust.

Since a large fraction of the massive star-formation at both low- and
high-$z$ has taken place in (U)LIRGs, their implied high
star-formation rates (SFRs) are expected to result in CCSN rates a
couple of orders of magnitude higher than in normal
galaxies. Therefore, a powerful tracer for starburst activity in
(U)LIRGs is the detection of CCSNe, since the SFR is directly related
to the CCSN rate.  However, most SNe occurring in ULIRGs are optically
obscured by large amounts of dust in the nuclear starburst
environment, and have therefore remained undiscovered by (optical) SN
searches.  Fortunately, it is possible to discover these CCSNe through
high-resolution radio observations, as radio emission is free from
extinction effects.  Furthermore, CCSNe are expected, as opposed to
thermonuclear SNe, to become strong radio emitters when the SN ejecta
interact with the circumstellar medium (CSM) that was expelled by the
progenitor star before its explosion as a supernova.  Therefore, if
(U)LIRGs are starburst-dominated, bright radio SNe are expected to
occur and, given its compactness and characteristical radio behaviour,
can be pinpointed with high-resolution, high-sensitivity radio
observations (e.g., SN 2000ft in NGC 7469 \cite{col01}; SN 2004ip in
IRAS 18293-3413, \cite{per07}; SN 2008cs in IRAS 17138-1017,
\cite{per08}, \cite{kankare08a}, \cite{kankare08b}; supernovae in Arp 299 \cite{neff04},
Arp 220 \cite{parra07} and Mrk 273 \cite{bon05}). However, due to 
(i) the large distances where (U)LIRGs are located and (ii) the 
likely contribution of a putative AGN, it is mandatory the
use of high-sensitivity, high-resolution radio observations to
disentangle the nuclear and stellar (mainly from young SNe)
contributions to the radio emission, thus probing the mechanisms
responsible for the heating of the dust in those regions.\\

\section{VLA imaging of local LIRGs}

\begin{figure}[t!]
\includegraphics[width=\textwidth]{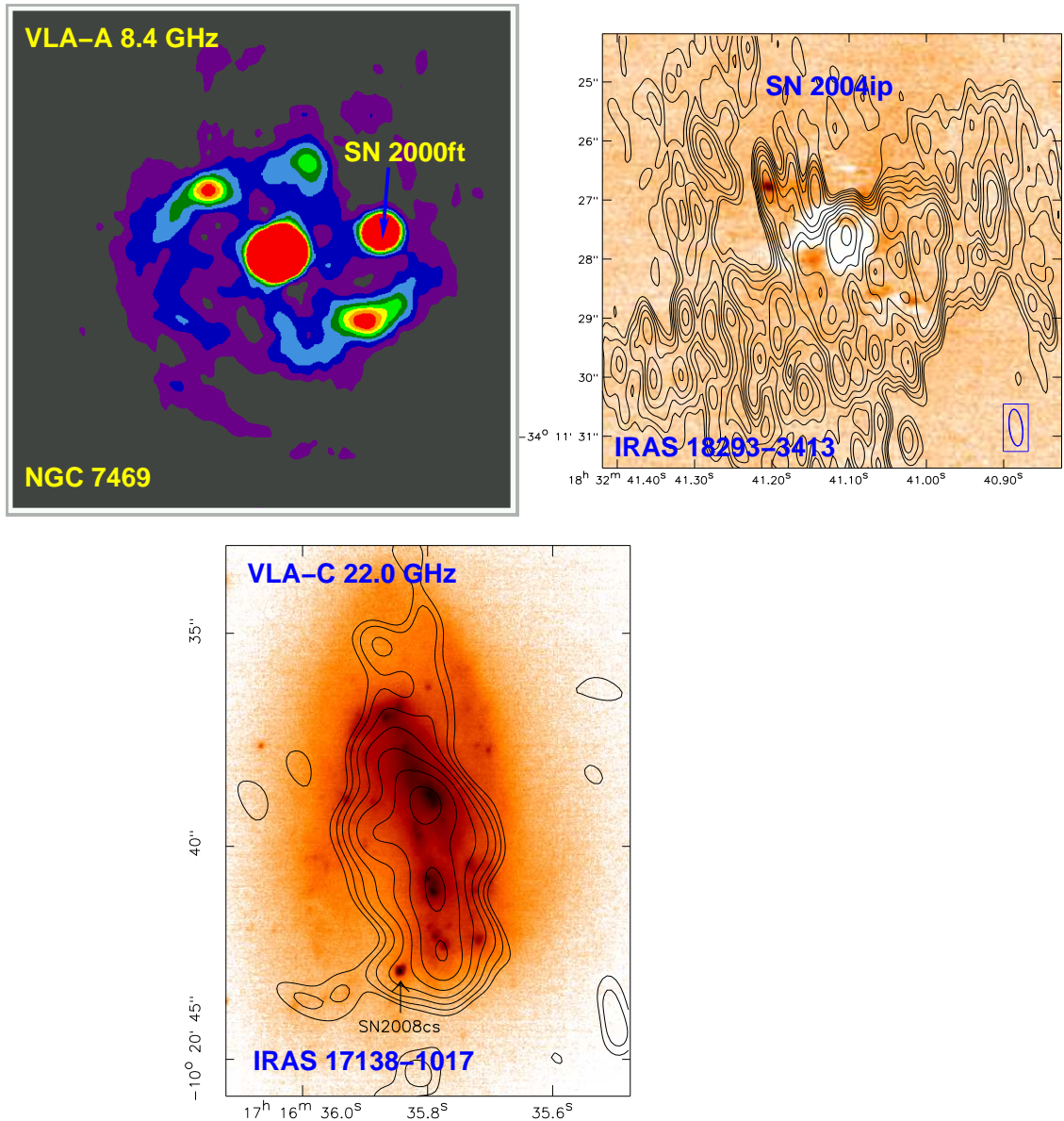}
\caption{ Top left: 8.4 GHz VLA discovery image of the radio supernova
  SN~2000ft in the galaxy NGC~7469 \cite{col01}. Top right: 8.4 GHz
  VLA contours of SN 2004ip, overlaid on top of the NIR 2.2$\mu$m
  discovery image of the supernova \cite{mattila07}, whose detection
  at radio wavelengths three years after its explosion confirmed it
  core-collapse nature \cite{per07}. 
  Bottom center: 22.0 GHz VLA contours of SN 2008cs, overlaid on top of a NIR 2.2$\mu$m
  GEMINI image \cite{per08}, \cite{kankare08a}, \cite{kankare08b}.
}
\label{fig,vla}
\end{figure}

The circumnuclear regions that are expected to host significant
starburst activity in nearby LIRGs are typically located at radial
distances of $\lsim$1~kpc. To unambiguouly detect the radio emission
from CCSNe, which highlight the presence of a recent starburst, it is
crucial that the radio observations be taken with high angular
resolution. In particular, the VLA-A at 8.4 GHz yields an angular
resolution of 0.3'', which is enough to discern the contribution of
exploding SNe from that of a putative AGN in nearby LIRGs 
(up to a distance of $\sim$100 Mpc).

In Figure \ref{fig,vla}, we show VLA images of several nearby ($D
\lsim$100 Mpc) LIRGs. In particular, we show VLA-A images of NGC~7469
($D=$70 Mpc), IRAS~18293-3413 ($D=$79 Mpc), and IRAS~17138-1017
($D=$75 Mpc), where a number of supernovae have been discovered
(SN~2000ft in NGC~7469 \cite{col01}); SN~2004ip in IRAS 18293-3413
\cite{per07}; SN 2008cs in IRAS 17138-1017 \cite{per08} and
\cite{kankare08a}, \cite{kankare08b}).  We note here that those SNe
were not discovered in the optical, since they are dust-enshrouded in
the dense medium of their host galaxies, but at radio wavelenghts (SN
2000ft) and in the NIR and radio (SN 2004ip and SN 2008cs).  Those
cases show that high-resolution radio observations of (circumn)nuclear
starbursts are crucial to unveil CCSNe in dust-enshrouded
environments.  In addition, if the exploding CCSNe are bright enough,
their monitoring allows to gain insight in both the progenitor and the
interaction between the supernova and its circumstellar medium (CSM),
and eventually the interstellar medium (ISM), by means of a radio
follow-up. For exmample, the monitoring of SN 2000ft showed that SN
2000ft is a type IIn SN \cite{alb06}, which result in long-lasting
radio events, e.g., SN 1986J in NGC 891 \cite{per02}, and seems to
follow an evolution similar to that of radio supernovae in normal
galaxies.

\begin{figure}[t!]
\includegraphics[width=\textwidth]{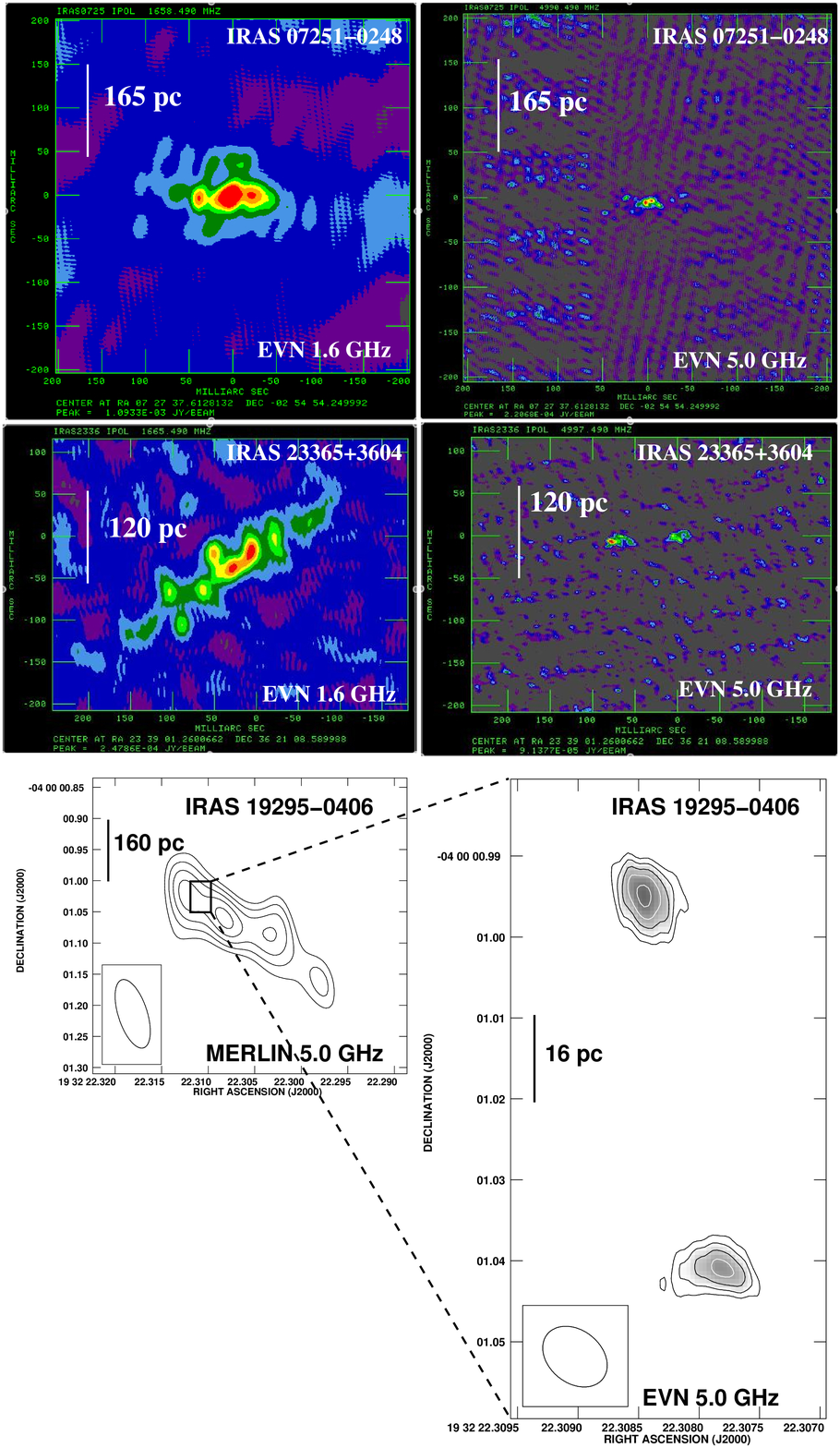}
\caption{Top panels: 1.6 and 5.0 GHz EVN images of IRAS 07251-0248.
Middle panels:  1.6 and 5.0 GHz EVN images of IRAS 2336+3604.
Bottom panels: 5.0 GHz MERLIN (left) and EVN (right) images 
of IRAS 19295-0406. (See text for details.)}
\label{fig,ulirgs}
\end{figure}

\section{EVN+MERLIN imaging of local ULIRGs}

We have recently embarked ourselves in a project to image at radio
wavelengths four of the most distant ULIRGs in the nearby universe,
combining simultaneous observations with the European VLBI Network
(EVN) and the Multi-Element Radio Linked Interferometer (MRLIN).  
The use of those arrays is mandatory because, at the distances where those
ULIRGs are located (from 250 Mpc up to 350 Mpc), the VLA 
neither is able to discern the compact radio emission from single supernovae, 
nor can separate it from a putative AGN contribution.

We have obtained the deepest and highest resolution radio images ever
of those ULIRGs, using quasi-simultaneous observations with the EVN at
1.6 and 5.0 GHz (see Figure \ref{fig,ulirgs} and \cite{cromero08}).
The top and middle panels show preliminary images of IRAS 07251-0248
(D=344 Mpc, L$_{\mathrm{FIR}}$ = 10$^{12.32}$\,L$_\odot $, CCSN rate
$\approx$ 8 SN/yr); IRAS 2336+3604 (D=252 Mpc, L$_{\mathrm{FIR}}$ =
10$^{12.13}$\,L$_\odot $, CCSN rate $\approx$ 5 SN/yr); IRAS
19295-0406 (D=338 Mpc, L$_{\mathrm{FIR}}$ = 10$^{12.37}$\,L$_\odot $,
CCSN rate $\approx$ 8 SN/yr); We have found a number of bright,
compact components, some of which are suggestive of CCSNe exploding in
the innermost regions of those ULIRGs, and whose radio luminosities
are typical of Type IIL and Type IIn SN, which yield bright radio SNe.
The images of IRAS 07251-0248 and IRAS 2336+3604 show that not all of
the the 5.0 GHz brightness peaks are coincident with those seen at 1.6
GHz.  As discussed in \cite{cromero08}, this is consistent with an
scenario where we are witnessing the radio emission of recently
exploding CCSNe, so that their 5.0 GHz emission would now be around
their peak, while their 1.6 GHz emission would still be rising. On the
other hand, we have also found several 1.6 GHz peaks without a clear
counterpart at 1.6 GHz. This can be explained if their emission arises
from CCSNe that are already in their optically thin phase, as
indicated by their two-point spectral indices.  Finally, the bottom
panels show the simultaneous 5.0 GHz MERLIN (left) and EVN (right)
images of IRAS 19295-0406.  Note that the MERLIN image cannot resolve
the radio emission from single (or chained of) supernovae. However,
the EVN can pinpoint the precise location of the compact radio
emissioin. As with the cases above discussed, the inferred
luminosities are in agreement with young, bright radio events, typical
of Type IIL and IIn supernovae. Future observations with the
EVN+MERLIN of those ULIRGs (already granted), will allow us to confirm
all of the SN candidates, as well as to precise locate the putative
AGN in those galaxies.

To summarize, high-resolution radio observations of (U)LIRGs in the
local universe are a powerful tool to probe the dominant dust heating
mechanism in their nuclear and circumnuclear regions, as they are free
from extinction, unlike optical observations.  The most direct way of
probing whether the main mechanism is due to recent starbursts is the
radio search of core-collapse supernovae (CCSNe), whose detection will
permit us to establish, independently of models, the CCSN rate for the
galaxies of our sample and, by assuming an Initial Mass Function
(IMF), their Star Formation Rates (SFR).  
%In addition, our
%observations will also allow us to constrain the nature of their SN
%progenitors, and will also help to understand the way these supernovae
%interact with their circumstellar medium in nearby (U)LIRGs.

\begin{small}
\paragraph*{Acknowledgments}
MAPT, AA, CRC, LC, and JMT acknowledge support by the Spanish
\textsl{Ministerio de Educación y Ciencia (MEC)} through grant AYA
2006-14986-C02-01.  MAPT is a Ram\'on y Cajal Post Doctoral Research
Fellow funded by the MEC and the \textsl{Consejo Superior de
  Investigaciones Científicas (CSIC)}.  MAPT, AA, CRC, and JMT also
acknowlegde support by the \textsl{Consejería de Innovación, Ciencia y
Empresa} of  \textsl{Junta de Andaluc\'{\i}a} through grants FQM-1747 and
TIC-126.  The European VLBI Network is a joint facility of European,
Chinese, South African and other radio astronomy institutes funded by
their national research councils.  The National Radio Astronomy
Observatory is a facility of the National Science Foundation operated
under cooperative agreement by Associated Universities, Inc.
\end{small}

\end{document}